\pdfoutput=1
\title{Reducing Complexity in Next-Generation \\ Multi-User MIMO Systems}
\author{
       \normalsize
        \textsc{Konstantinos Ntougias}
            \qquad
        \normalsize
        \textsc{Dimitrios Ntaikos}
        \qquad
         \normalsize
        \textsc{Constantinos B. Papadias}
            \qquad
        \mbox{}\\ %
        Athens Information Technology (AIT), 44 Kifissias Avenue, 15125 Maroussi, Greece.\\
        \mbox{}\\ %
        \normalsize
            \texttt{kontou}
        \textbar{}
            \texttt{dint}
             \textbar{}
        \normalsize
            \texttt{cpap}
         \normalsize
            \texttt{@ait.gr}
}

\date{}

\documentclass[twocolumn,a4paper, final]{article}
\usepackage{graphicx}
\usepackage{subfig}
\usepackage{multirow}
\usepackage{array}
\usepackage[noadjust]{cite}
\usepackage{url}
\usepackage{diagbox}
\usepackage{amsmath}
\usepackage{amssymb}
\usepackage{bm}
\usepackage[margin=1in]{geometry}

\newcommand{\matr}[1]{\mathbf{#1}}
\providecommand{\keywords}[1]{\textbf{\textit{Index terms---}} #1}

\begin{document}
\maketitle
\begin{abstract}
Recently, several advanced multi-antenna radio communications technologies have emerged to meet the increased capacity demands in wireless multi-user networks.  Despite their great potential, the extent of these techniques' practical applicability still remains questionable, since they have to face either backhaul limitations or cost and hardware constraints. In this paper, we propose a new system solution which includes network architecture, antenna technology and radio transmission protocol to reduce drastically the hardware complexity and cost as well as the channel state information / user data feedback requirements of multi-user multi-antenna wireless networks. We focus on the forward link of an interference channel in a cloud radio access network setup wherein an arbitrary number of remote radio heads are each equipped with a single radio frequency module parasitic antenna array and wish to send data to their respective single-antenna user terminals, while co-existing in time and frequency. Base stations select cooperatively the optimal combination of pre-determined beams prior transmission. Our proposed approach is able to achieve the aforementioned goals, while offering significant downlink sum-rate gains due to the available spatial degrees of freedom. 
\end{abstract}


\section{Introduction}\label{sec:1}
Over the last decade, an enormous growth of mobile data traffic has been witnessed. As several studies indicate~\cite{TRAFFIC}\cite{CISCO}, this trend is likely to continue in the years to come in an even more abrupt pace. In view of the well-known scarcity and high cost of the radio spectrum, current cellular mobile broadband networks incorporate multiple-input multiple-output (MIMO) technology as a response to the increased capacity demands.

Traditional single-user MIMO (SU-MIMO) focuses on the physical layer performance of the communication between a base station (BS) and a mobile terminal (MT) over a point-to-point link. It exploits the additional degrees of freedom (DoFs) provided by the use of multiple antennas to enhance spectral efficiency through spatial multiplexing of individual data streams or to reduce co-channel interference by spatially focusing transmissions and separating co-channel signals.

Multi-user MIMO (MU-MIMO) is an evolution of SU-MIMO that enhances link-level performance by exploiting DoFs to spatially separate users, thus enabling a BS to communicate simultaneously with multiple mobile users (MUs) over a multipoint-to-point (uplink) or a point-to-multipoint (downlink) channel.

Recently, new MIMO paradigms, which aim at improving system-level performance, have been introduced. Their common characteristic is that they increase significantly the available DoFs to boost the performance of the system. More specifically, cooperative MIMO~\cite{COMP}\cite{CoMP_Clerckx} enables the cooperation between different BSs to mitigate inter-cell interference and increase area spectral efficiency and system capacity, while massive MIMO~\cite{MASSIVE}\cite{Marzetta} makes use of an excessive number of antennas at the cell site to orthogonalize MUs in the spatial domain.

It is expected that both these technologies will be an integral part of future fifth generation (5G) systems. However, their transition from theory into implementation requires the addressing of some important practical issues.

Cooperative MIMO refers to a collection of techniques with varying level of cooperation, from network MIMO which requires the sharing of user data and global channel state information (CSI) between the individual BSs to cooperative beamforming which requires only the exchange of precoding matrices between the cooperating nodes. In practice, backhaul latency limits the effectiveness of these techniques, while backhaul capacity constraints reduce the achievable level of cooperation and the corresponding performance gains.

Massive MIMO, on the other hand, places a heavy burden, in terms of cost, to the mobile network operators due to the large number of radio-frequency (RF) chains (one per active antenna element) that is required for its implementation. Moreover, the tight spacing of antenna elements may lead to reduced efficiency due to spatial correlations.

In addition, as the number of BSs or/and antennas increases, channel estimation becomes a challenging task. 

In this paper, we describe a system setup that is promising in its ability to address these challenges. We focus on the study of a radio transmission protocol that takes advantage of the attributes of the employed technologies to achieve substantial performance gains, while reducing significantly system complexity.

The structure of the remainder of the paper is as follows: In Section~\ref{sec:2}, the considered system architecture and antenna technology is presented. Next, the system and channel model is described in Section~\ref{sec:3}. We continue in Section~\ref{sec:4} with a presentation of the proposed radio protocol.  In Section~\ref{sec:5} we present numerical performance results for different flavors of the proposed transmission scheme and compare them against performance bounds. We then discuss in Section~\ref{sec:6} about the complexity reduction that is achieved with the considered radio protocol as well as about the relevant complexity-feedback overhead and complexity-performance tradeoffs. In Section~\ref{sec:7} we study the performance of the transmission scheme when CSI is imperfect. Finally, in Section~\ref{sec:8} we present our conclusions and discuss about future extensions of our work. 

\section{System Architecture and Technologies}\label{sec:2}
According to the discussion in Section~\ref{sec:1}, we define three main criteria that are important to be met in order to enable the use of Cooperative / Massive MIMO in practice:
\begin{enumerate}
\item{Low-cost implementation of systems with many antennas.}
\item{Low-delay communication between cooperating nodes.}
\item{Minimization of the information that has to be exchanged between the cooperating base stations.}
\end{enumerate}

In order to fulfill these requirements, all aspects of system design have to be taken into account, that is, network architecture, antenna technologies, and radio transmission protocols. In this Section, we describe two proposed components that can meet the first two of these objectives. Next, we will focus on a proposed radio scheme that can be employed by this system to reduce feedback complexity.

\begin{figure*}[!t]
\centering
\includegraphics[scale = 0.5]{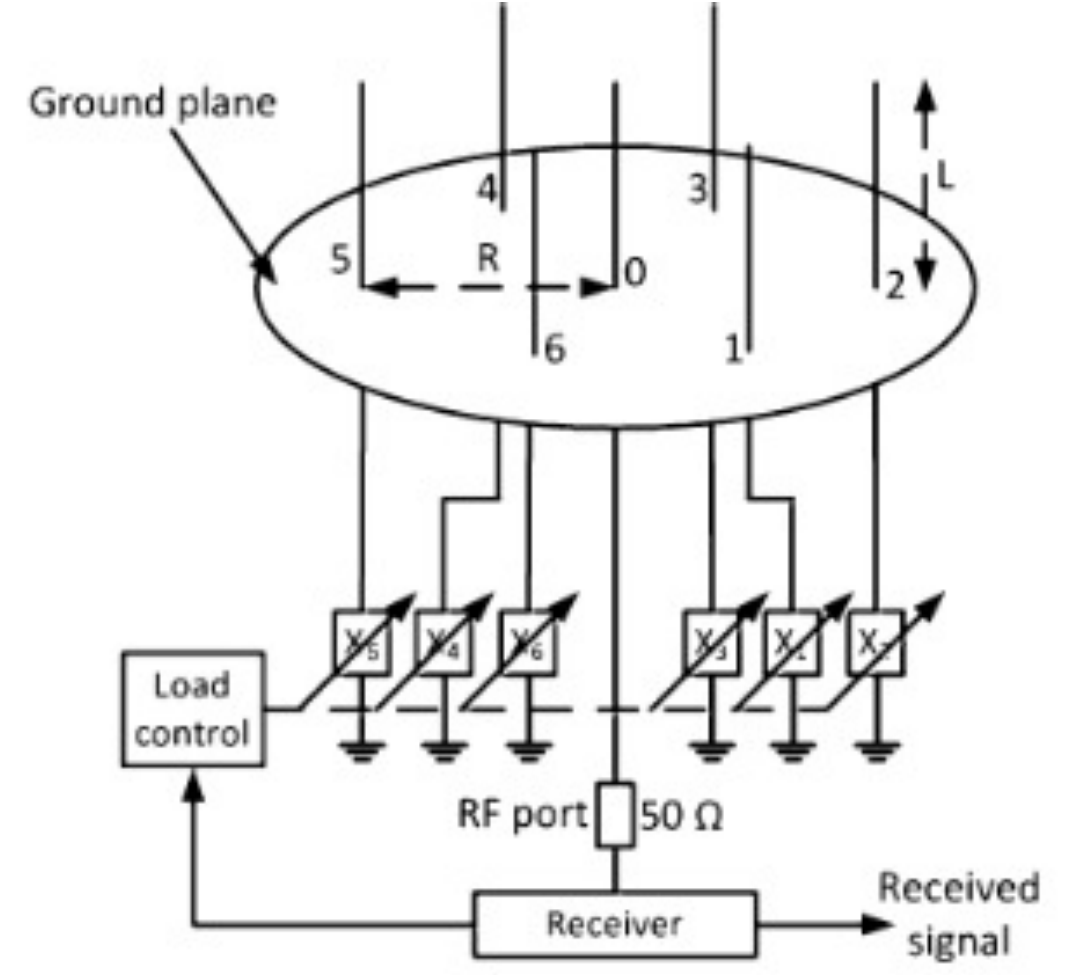}
\caption{The ESPAR paradigm for beam shaping in the analog domain with digital control~\cite{ESPAR}.}
\label{fig:1}
\end{figure*}

\subsection{ESPAR Antennas}\label{subsec:2.1}
Advanced parasitic antenna arrays, often called electronically steerable parasitic array radiators (ESPARs), provide multi-antenna functionality using fewer active elements (and, as a consequence, fewer RF units) than conventional antenna arrays. They accomplish this by adjusting the loads of parasitic elements that are placed in the vicinity of the active element(s), thus controlling the corresponding induced currents caused by mutual coupling to form and steer beams ~\cite{ESPAR}. This concept is illustrated in Fig.~\ref{fig:1}.

ESPAR technology can be viewed as an attractive enabler of massive MIMO, since it allows the implementation of large antenna systems with much reduced complexity, size and cost.

\subsection{Cloud-RAN/RRH}\label{subsec:2.2}
Cloud Radio Access Network (Cloud-RAN or C-RAN)~\cite{CRAN}\cite{HARP}  is a split base station architecture which separates the baseband unit (BBU) from the radio unit. More specifically, in C-RAN, remote radio units (RRUs), also called remote radio heads (RRHs), remain at the cell site, while BBUs are centralized and virtualized using cloud technology. Baseband and radio resources are allocated to ``virtual base stations'' in real time according to processing and radio coverage needs. Remote radio heads are connected with the centralized BBU pool through optical fibers. Protocols such as Common Public Radio Interface (CPRI)~\cite{CPRI} are used to enable the communication between the virtualized baseband pool and the RRHs over this new network segment, which is often referred to as the mobile fronthaul.

C-RAN is hence considered to be a compelling candidate technology for bringing cooperative MIMO communication into reality, due to the centralization and virtualization of the baseband and the use of optical transmission technologies at the fronthaul which have the potential to meet the stringent delay and bandwidth requirements of these advanced techniques~\cite{Aleksandra}.

\section{System and Channel Model}\label{sec:3}
In this Section, we introduce the system and channel model of our setup. Before getting in the details, let us present the notation that we follow throughout the paper.

\emph{Notation}: With $a$, $\matr{a}$, and $\matr{A}$ we denote a (in general, complex-valued) scalar, vector, and matrix, respectively. $\matr{A}_{i,j}$ represents the ($i,j$) (row, column) element of $\matr{A}$. The Hermitian (transpose conjugate), determinant, and rank of $\matr{A}$ are denoted by $\matr{A}^\dagger$, $\operatorname{det}(\matr{A})$ and $\textrm{rank}(\matr{A})$, respectively, while the inverse and the Moore-Penrose pseudo-inverse of $\matr{A}$ are denoted by $\matr{A}^{-1}$ and $\matr{A}^{+}$, respectively. $\textrm{Tr}(\matr{A})$ represents the trace of $\matr{A}$ and $\|\matr{A}\|_{F}$ denotes its Frobenius norm, whereas $\textrm{diag}(\matr{a})$ represents a diagonal matrix with the elements of vector $\matr{a}$ at its main diagonal. $\|\matr{a}\|$ stands for the Euclidean norm of vector $\matr{a}$. $\matr{0}_{n}$ and $\matr{I}_{n}$ denote the $n \times n$ zero and identity matrix, respectively (the subscript indicating the dimension of the matrix may be omitted whenever is irrelevant). $\mathbb{C}$ denotes the set of complex numbers, $\mathbb{E}\{\cdot\}$ represents the expectation operator,  $X \sim \mathcal{CN}\left(0,\sigma^{2}\right)$ refers to a complex-valued random variable (RV) $X$ following the Gaussian distribution with zero mean (i.e., $\mathbb{E}\{X\} = 0$) and variance $\sigma_{x}^{2}$, and $\matr{a} \sim \mathcal{CN}\left(\matr{0}, \sigma_{a}^{2}\matr{I}\right)$ represents a complex-valued Gaussian vector with mean matrix $\mathbb{E}\{\matr{a}\} = \matr{0}$ and covariance matrix $\matr{R}_{aa} = \mathbb{E}\left\{\matr{a}\matr{a}^{\dagger}\right\} = \sigma_{a}^{2}\matr{I}$, i.e., the elements of $\matr{a}$ are complex-valued RVs with zero mean and variance $\sigma_{a}^{2}$. 
       
 \begin{figure*}[!t]
\centering
\includegraphics[scale = 0.5]{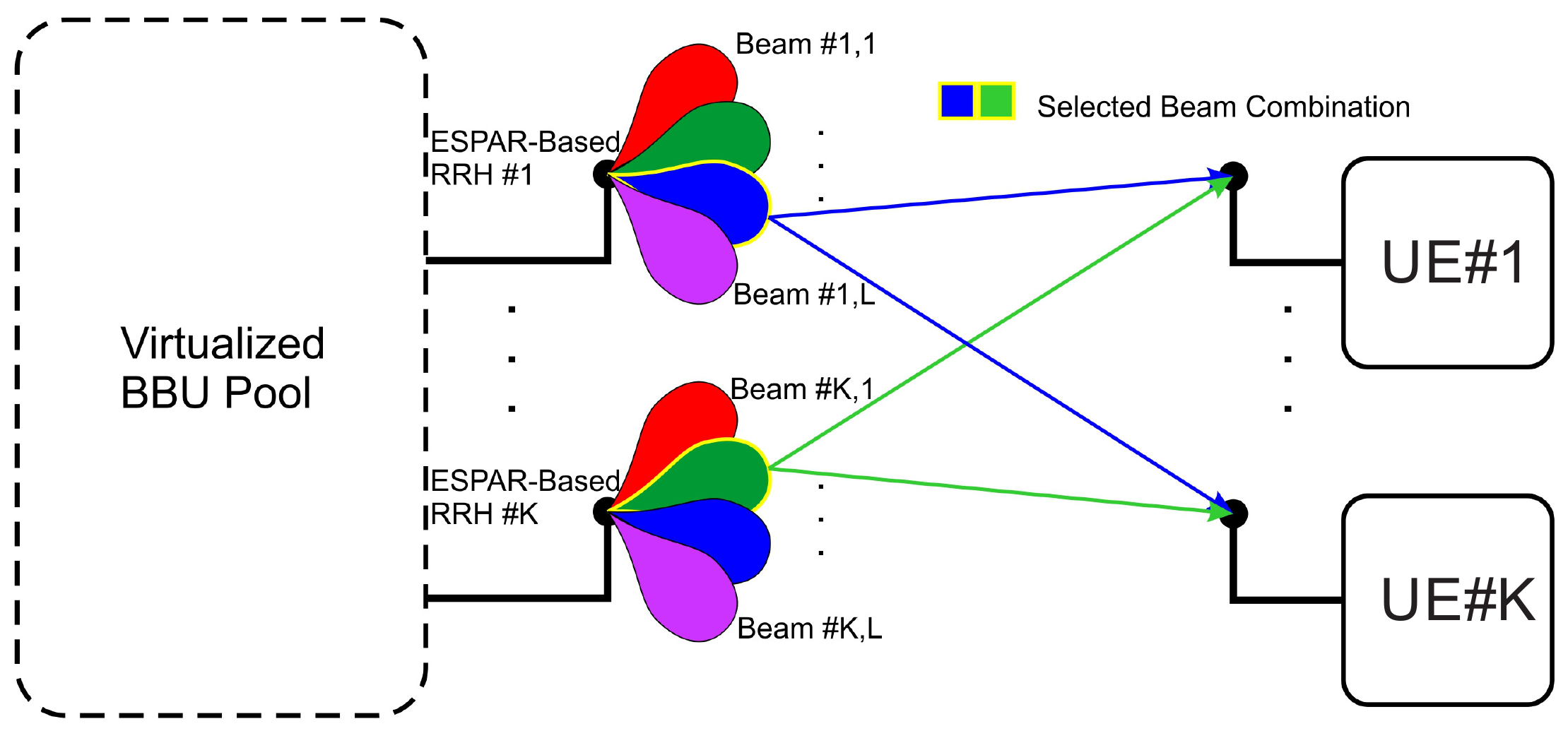}
\caption{System model. Note the transformation of the channel to a ``beam-domain'' interference / broadcast channel.}
\label{fig:2}
\end{figure*}

\subsection{System Model}\label{subsec:3.1} 
We consider the downlink of a C-RAN system with $K$ RRHs, each representing an individual BS and being equipped with a single-RF ESPAR antenna, where each one of them wishes to communicate with a user equipment (UE) having a single antenna. All transmissions take place simultaneously and over the same frequency band. 

Note that due to the C-RAN architecture, the BSs can efficiently share information. The MUs, on the other hand, do not cooperate.

Each ESPAR antenna is able to generate $L$ distinct predetermined beams. According to some criterion, the best $K$-tuple of beams is selected for transmission from the $L^{K}$ different beam combinations in total, with each one of these beams being generated at a different RRH, as seen in Fig.~\ref{fig:2}. (The beam selection criteria are presented in Section~\ref{sec:4}.)

The corresponding ``beam-domain transformed'' channel between these $K$ transmitter-receiver (TX-RX) pairs is equivalent to a single-input single-output (SISO) $K$-user interference channel (IFC), since each TX has a single active antenna. However, we should not ignore the fact that the TXs are equipped with parasitic antenna arrays and make use of beamforming, in contrast to conventional SISO TXs which utilize omni-directional antennas.   

Mathematically, this SISO IFC with $K$ TXs and $K$ RXs having $\left\{N_{T,k}\right\}_{k=1}^{K} = 1$ and $\left\{N_{R,k}\right\}_{k=1}^{K} = 1$ active antennas each, respectively, can be viewed as a multi-user multiple-input single-output (MU-MISO) $K$-user IFC formed by a composite TX with $N_{T} = \sum_{k=1}^{K}N_{T,k} = K$ antennas and $K$ RXs, each with $\left\{N_{R,k}\right\}_{k=1}^{K} = 1$ antenna.

We assume narrowband transmission (i.e., flat-fading channel), such that no inter-symbol interference is caused by multipath. This is the case, for example, in indoor environments with relatively small delay spread as well as when orthogonal frequency division multiplexing (OFDM) transmission is employed to convert a frequency-selective channel into parallel frequency-flat sub-channels. We also assume block-fading, such that the channel remains fixed during the transmission of each symbol. Therefore, we can omit time dependence in the system model.  

Under these assumptions, the baseband received signal at user $k$, where $k = 1, \dots, K$, \emph{for a selected beam combination}, can be expressed as
\begin{equation}\label{eq:1}
y_{k} = \sum_{m=1}^{K}h_{k,m}s_{m} + n_{k},\quad k = 1, \dots, K
\end{equation}
where $s_{m}\in\mathbb{C}$ is the complex symbol transmitted by $\textrm{Tx}_{m}$, $y_{k}\in\mathbb{C}$ is the complex received symbol at $\textrm{Rx}_{k}$, $n_{k} \sim \mathcal{CN}(0,\sigma_{n}^{2})$ is complex additive Gaussian noise with zero mean and variance $\sigma_{n}^{2}$, and $h_{k,m}$ is the complex channel coefficient from $\textrm{Tx}_{m}$ to $\textrm{Rx}_{k}$.

Equivalently, we can write
\begin{equation}\label{eq:2}
y_{k} = \matr{h}_{k}^{\dagger}\matr{s} + n_{k},\quad k = 1, \dots, K
\end{equation}
where $\matr{h}_{k}\in\mathbb{C}^{K \times 1}$ is the vector of the channel coefficients between $\textrm{Rx}_{k}$ and $\{\textrm{Tx}_{m}\}_{m=1}^{K}$ and $\matr{s}\in\mathbb{C}^{K \times 1}$ is the vector of transmitted symbols. 

The input-output model of the overall channel is
\begin{equation}\label{eq:3}
\matr{y} = \matr{H}\matr{s} + \matr{n},
\end{equation}
where $\matr{y}\in\mathbb{C}^{K \times 1}$ is the vector of received symbols; $\matr{n}\in\mathbb{C}^{K \times 1}$ is a complex-valued Gaussian noise vector with $\matr{n} \sim \mathcal{CN}(\matr{0}, \sigma_{n}^{2}\matr{I}_{K})$, i.e., with mean matrix $\mathbb{E}\{\matr{n}\} = \matr{0}_{K}$ and covariance matrix $\matr{R}_{nn} = \mathbb{E}\left\{\matr{n}\matr{n}^{\dagger}\right\}= \sigma_{n}^{2}\matr{I}_{K}$; and $\matr{H}\in\mathbb{C}^{K \times K}$ is the channel matrix expressed as
\begin{equation}\label{eq:4}
\matr{H} = \begin{bmatrix} \matr{h}_{1}, & \dots, & \matr{h}_{K} \end{bmatrix}^{\dagger}.
\end{equation}
\begin{figure*}[!t]
\centering
\includegraphics[scale = 0.4]{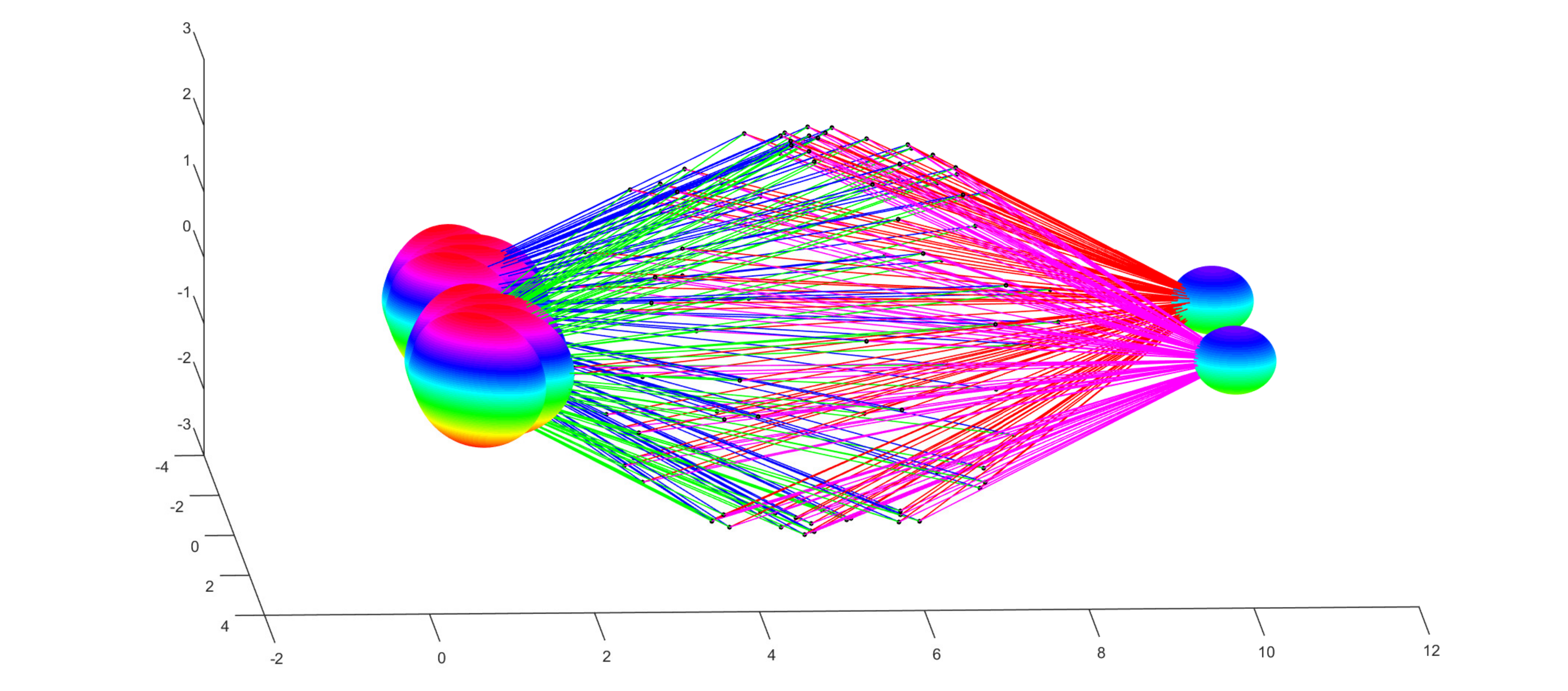}
\caption{Single-bounce scattering model with 100 scatterers for a system with $K = 2$ and $L = 4$ beams per transmitter.}
\label{fig:3}
\end{figure*}
The total transmitted power is constrained to $P$, i.e., 
\begin{equation}\label{eq:5}
\textrm{Tr}\left(\matr{R}_{ss}\right) = \textrm{Tr}\left(\mathbb{E}\left\{\matr{s}\matr{s}^{\dagger}\right\}\right) \leq P,
\end{equation}
where $\matr{R}_{ss} = \mathbb{E}\left\{\matr{s}\matr{s}^{\dagger}\right\}$ is the covariance matrix of the transmitted signal vector $\matr{s}$.

For convenience, and without loss of generality, we assume that all transmitted symbols $\left\{s_{k}\right\}_{k=1}^{K}$ have equal variance $\sigma_{s}^{2} = P/K = 1$ such that $\matr{R}_{ss} = \sigma_{s}^{2}\matr{I}_{K} = \matr{I}_{K}$ and $P = \textrm{Tr}\left(\matr{R}_{ss}\right) = K$.

\subsection{Channel Model}\label{subsec:3.2} 
In order to capture the effect of the beam radiation patterns, we consider, without loss of generality, a geometry-based single-bounce scattering statistical model~\cite{SINGLE}\cite{MIMO}. In the single-bounce approach, each transmit/receive path is broken into two sub-paths: transmitter-to-scatterer and scatterer-to-receiver (described by their direction of departure, direction of arrival, and path distance). The scatterer itself is modeled typically via the introduction of a random attenuation and phase shift. In this paper, we assume that the scatterers are randomly distributed on the surface of a sphere that is placed in the middle of the distance between the TXs and the RXs, as seen in Fig.~\ref{fig:3}. This model has been selected because, while it is relatively simple in comparison with other models, it captures small-scale fading and shadowing effects sufficiently well.

In order to facilitate performance analysis, we normalize the channel matrix. More specifically, for fixed $\matr{H}$ (e.g., when a given channel realization over the coherence time of the channel is considered), the normalized channel matrix $\overset{\sim}{\matr{H}}$ is given by~\cite{NORM}:
\begin{equation}\label{eq:6}
\overset{\sim}{\matr{H}} = a\matr{H},
\end{equation}
 where, in the approach that we followed, the normalization constant is expressed as
\begin{equation}\label{eq:7}
|a|^{2} = \frac{N_{T}N_{R}}{\left\|\matr{H}\right\|_{F}^{2}} = \frac{K^{2}}{\left\|\matr{H}\right\|_{F}^{2}},
\end{equation}
such that the power gain of the channel is
\begin{equation}\label{eq:8}
\textrm{Tr}\left(\overset{\sim}{\matr{H}}\overset{\sim}{\matr{H}}^{\dagger}\right) = \left\|\overset{\sim}{\matr{H}}\right\|_{F}^{2} = N_{T}N_{R}= K^{2}.
\end{equation}

Of course, when the channel coefficients are RVs (e.g., longer time intervals are of interest), the expectation over $\matr{H}$ should be taken in the appropriate expressions.

Note that in Eqs.~\eqref{eq:1}--\eqref{eq:5}, as well as throughout the paper, we denote the normalized channel matrix simply as $\matr{H}$, for convenience.

\section{Radio Transmission Protocol}\label{sec:4}
System operation is divided into three phases, namely, the learning phase, the selection phase, and the transmission phase. In the learning phase, each UE acquires information about the quality or the coefficients of each channel formed by the $L^{K}$ different beam combinations at the TXs. In the selection phase, the BSs select jointly the best $K$-tuple of beams, according to the information that has been fed back to them by the UEs. Finally, in the transmission phase, communication takes place. The transmission strategy depends on the channel information available at the TXs.

\subsection{Learning Phase}\label{subsec:4.1}
There exist two different types of channel information that a UE may send back to the corresponding BS:
\begin{enumerate}
\item{Its signal to interference plus noise ratio (SINR) for each beam combination at the TXs.}
\item{The CSI for the resulting ``beam-domain transformed channel'', i.e., the complex-valued coefficients for the self- and cross-channels of the total $K \times K$ channel.}
\end{enumerate}

\subsection{Beam Selection Phase}\label{subsec:4.2}
According to the type of channel information that the BSs acquired through feedback, the following beam selection criteria apply:
\\\\
\hspace*{3.2mm}\textbf{Beam Selection Rule \#1:} Select the $K$-tuple of beams that results in the largest sum of SINRs, i.e., the beam combination that maximizes $\sum_{k=1}^{K}\gamma_{k}$.
\\\\
\hspace*{3.2mm}\textbf{Beam Selection Rule \#2:} Select the $K$-tuple of beams that results in the equivalent channel matrix with the largest product $\matr{H}\matr{H}^{\dagger}$.
\\

We should note that, even though our goal in this work is to generate beams using parasitic antenna arrays, since ESPAR technology reduces the cost, size, and complexity of the system, the general principle holds also for beams generated by conventional antenna arrays.

\subsection{Transmission Phase}\label{subsec:4.3}
In this phase, the transmission strategy that will be followed to send information symbols over the selected beams is determined. The performance metric that we wish to increase as much as possible while maintaining complexity and CSI feedback overhead low is the sum-rate throughput. 

If  beam selection is based on the measured SINRs at the RXs (i.e., the beams that maximize the sum of the received SINRs have been selected), then there is no additional processing taking place at the TXs. The sum-rate in this non-precoded transmission case is given by
\begin{equation}\label{eq:9}
R_{\textrm{NP}} = \sum_{k=1}^{K}R_{k}^{(\textrm{NP})} = \sum_{k=1}^{K}\log_{2}\left(1+ \gamma_{k}^{(\textrm{NP})}\right) ,
\end{equation}
where
\begin{equation}\label{eq:10}
 R_{k}^{(\textrm{NP})} = \log_{2}\left(1+ \gamma_{k}^{(\textrm{NP})}\right)
\end{equation}
is the rate achieved by user $k$ and $\gamma_{k}$ is the SINR of user $k$ which is expressed as
\begin{align}
\gamma_{k}^{(\textrm{NP})} &= \frac{\left|h_{k,k}\right|^{2}\sigma_{k}^2}{\sum_{m \neq k}\left|{h}_{k,m}\right|^{2}\sigma_{m}^{2} + \sigma_{n}^{2}} \nonumber \\ 
&= \frac{\left|h_{k,k}\right|^{2}}{\sum_{m \neq k}\left|{h}_{k,m}\right|^{2} + \sigma_{n}^{2}}. \label{eq:11}
\end{align}
since $\left\{\sigma_{k}^{2}\right\}_{k=1}^{K} = \sigma_{s}^{2} = 1$.

If, on the other hand, the channel matrices have been fed back to the TXs (i.e., the beams that result in an equivalent channel with the largest product $\matr{H}\matr{H}^{\dagger}$ have been selected), then the TXs jointly precode the transmit vector $\matr{s}$, thus transforming the MU-MISO IFC into a MU-MISO broadcast channel (BC). 

We assume that linear precoding is employed by the system, in order to avoid the high RX complexity required by more advanced transmission techniques~\cite{MIMO}. Linear precoding includes a family of simple but sub-optimal pre-processing techniques that exploit CSI at the transmitter (CSIT) to improve MU-MIMO performance--e.g., to increase the sum-rate throughput or to minimize the aggregated bit error rate. (Unless stated otherwise, perfect CSIT is assumed.) 

A linear precoder generates a precoded signal vector as a linear transformation of the original symbol vector:
\begin{equation}\label{eq:12}
\matr{s}^{\prime} = \matr{W}\matr{s}.
\end{equation}

Hence, the received signal vector is expressed as
\begin{equation}\label{eq:13}
\matr{y} = \matr{H}\matr{s}^{\prime} + \matr{n}.
\end{equation}
That is,
\begin{equation}\label{eq:14}
\underset{[K \times 1]}{\matr{y}} = \underset{[K \times K]}{\matr{H}}\underset{[K \times K]}{\matr{W}}\underset{[K \times 1]}{\matr{s}} + \underset{[K \times 1]}{\matr{n}}. 
\end{equation}

Equivalently, we can write
\begin{equation}\label{eq:15}
y_{k} = \matr{h}_{k}^{\dagger}\matr{w}_{k}s_{k} + \sum_{\underset{m\neq k}{m=1}}^{K}\matr{h}_{k}^{\dagger}\matr{w}_{m}s_{m} + n_{k},
\end{equation}
where $\matr{h}_{k}\in\mathbb{C}^{K \times 1}$, $\matr{w}_{k}\in\mathbb{C}^{K \times 1}$ and $s_{k}$ are the channel vector, precoding vector, and data stream of user $k$, respectively; $\matr{H}$ is the channel matrix; and $\matr{W}$ is the precoding matrix expressed as
\begin{equation}\label{eq:16}
\matr{W} = \begin{bmatrix} \matr{w}_{1} & \cdots & \matr{w}_{K}\end{bmatrix}.
\end{equation}

Note that the second term at the right-hand side of Eq.~\eqref{eq:15} represents the multi-user interference (MUI).

The SINR of user $k$ for this linear precoding scheme is given by
\begin{equation}\label{eq:17}
\gamma_{k}^{(\textrm{LP})} = \frac{\left\|\matr{h}_{k}^{\dagger}\matr{w}_{k}\right\|^{2}}{\sum_{m\neq k}\left\|\matr{h}_{k}^{\dagger}\matr{w}_{m}\right\|^{2} + \sigma_{n}^{2}}.
\end{equation}
The expressions of the $k$th user's rate $R_{k}^{(\textrm{LP})}$ and the sum-rate of the system $R_{\textrm{LP}}$ are similar with the ones given in Eq.~\eqref{eq:10} and~\eqref{eq:9}, respectively, for the non-precoded system.  

The Tx power constraint is
\begin{align}
\textrm{Tr}\left(\mathbb{E}\left\{\matr{s}^{\prime}\left(\matr{s}^{\prime}\right)^{\dagger}\right\}\right) &= \textrm{Tr}\left(\mathbb{E}\left\{\left(\matr{W}\matr{s}\right)\left(\matr{W}\matr{s}\right)^{\dagger}\right\}\right) \nonumber \\
&= \textrm{Tr}\left(\mathbb{E}\left\{\matr{W}\matr{s}\matr{s}^{\dagger}\matr{W}^{\dagger}\right\}\right) \nonumber \\
&= \textrm{Tr}\left(\matr{W}\mathbb{E}\left\{\matr{s}\matr{s}^{\dagger}\right\}\matr{W}^{\dagger}\right) \nonumber \\
&= \textrm{Tr}\left(\matr{W}\matr{R}_{ss}\matr{W}^{\dagger}\right)\leq P. \label{eq:18}
\end{align}

We propose the use of zero-forcing (ZF) precoding, which provides a promising compromise between complexity and performance~\cite{ZF}, especially at the high signal to noise ratio (SNR) regime~\cite{Goldsmith}.

The reasoning behind ZF precoding is to employ a linear transformation to the transmit signal vector, according to Eq.~\eqref{eq:12}, such that MUI becomes null, i.e., so that each user $k = 1, \dots, K$ receives no interference from the signals intended for the other users~\cite{MIMO}:
\begin{equation}\label{eq:19}
\left\|\matr{h}_{k}^{\dagger}\matr{w}_{m}^{(\textrm{ZF})}\right\| = 0,\quad m\neq k
\end{equation}
Interference suppression at the TX is important in this setup, since single-antenna, non-cooperating RXs are unable to eliminate interference--the best they can do is to treat it as noise.

A common approach to accomplish this goal involves the inversion of the channel matrix at the TX in order to create orthogonal channels between the TX and the RXs (i.e., to diagonalize the effective channel). Channel inversion implies, under the flat-fading assumption, the multiplication of the original transmit vector signal with the right Moore-Penrose pseudo-inverse of the channel matrix $\matr{H}$, such that the matrix of the composite channel is the identity matrix. That is,
\begin{equation}\label{eq:20}
\matr{W}_{\textrm{ZF}} = \matr{H}^{+} \Rightarrow \begin{matrix} \begin{aligned}\matr{s}^{\prime} &= \matr{W}_{\textrm{ZF}}\matr{s} = \matr{H}^{+}\matr{s} \\ \matr{H}\matr{W}_{\textrm{ZF}} &= \matr{H}\matr{H}^{+} = \matr{I} \end{aligned} \end{matrix}
\end{equation}
where
\begin{equation}\label{eq:21}
\matr{H}^{+} = \matr{H}^{\dagger}\left(\matr{H}\matr{H}^{\dagger}\right)^{-1}.
\end{equation}

In practice, though, we typically normalize the precoding matrix $\matr{W}_{ZF}$ in order to set the transmit power (after precoding) to a fixed value, independent of the channel $\matr{H}$, according to the given power constraint.  An often used normalization, which is referred to as equal receive power (ERP) normalization, takes the following form~\cite{MATLAB}:
\begin{equation}\label{eq:22}
\matr{W}_{\textrm{ERP}} = \sqrt{\beta}\matr{H}^{+} = \sqrt{\beta}\matr{H}^{\dagger}\left(\matr{H}\matr{H}^{\dagger}\right)^{-1}.
\end{equation}
Under our assumption $\sigma_{s}^{2} = 1 \Rightarrow \mathbb{E}\left\{\matr{s}\matr{s}^{\dagger}\right\} = \matr{R}_{ss} = \matr{I}_{K} \Rightarrow \textrm{Tr}\left(\matr{R}_{ss}\right) = K$ and by forcing the total transmit power after precoding to remain equal to $P = K$, that is, 
\begin{equation}\label{eq:23}
\textrm{Tr}\left(\matr{W}_{\textrm{ERP}}\matr{R}_{ss}\matr{W}_{\textrm{ERP}}^{\dagger}\right) = K, 
\end{equation}
the power normalization factor $\beta$ is given by~\cite{Beta}\begin{subequations}\label{eq:24}
\begin{align}
\beta &= \frac{1}{\textrm{Tr}\left(\left(\matr{H}\matr{H}^{\dagger}\right)^{-1}\right)} \label{eq:24a}\\
&= \frac{1}{\left\|\left(\matr{H}\matr{H}^{\dagger}\right)^{-1}\right\|_{F}^{2}} \label{eq:24b}\\
&=  \frac{1}{\sum_{k=1}^{K}1/\left(\lambda_{k}^{2}\right)} \label{eq:24c}
\end{align}
\end{subequations}  
where $\lambda_{k}$ is the $k$th singular value of $\matr{H}$.

When the channel matrix is square, as it is in our case where $N_{T}=K$, then the precoding matrix is simply given by~\cite{Square}:
\begin{equation}\label{eq:25}
\matr{W}_{\textrm{ERP}} = \sqrt{\beta}\matr{H}^{-1}.
\end{equation}

In any case, the corresponding received signal vector is given by:
\begin{align}
\matr{y} &= \matr{H}\matr{s}^{\prime} + \matr{n} \nonumber \\
&= \matr{H}\left(\matr{W}_{\textrm{ERP}}\matr{s}\right) + \matr{n} \nonumber \\
&= \left(\matr{H}\matr{W}_{\textrm{ERP}}\right)\matr{s} + \matr{n} \nonumber \\
&= \sqrt{\beta}\matr{s} + \matr{n}. \label{eq:26}
\end{align}

Note that the scaling factor in Eq.~\eqref{eq:26} may lead in reduced SNR at the receiver if the channel matrix is ill-conditioned, i.e., if one of its singular values is very large.

The SINR at user $k$ is given by Eq.~\eqref{eq:17} by setting the MUI at the denominator equal to zero and noting that $\matr{H}\matr{W}_{\textrm{ERP}} = \sqrt{\beta}\matr{I}$:
\begin{equation}\label{eq:27}
\gamma_{k}^{(\textrm{ERP})} = \frac{\beta}{\sigma_{n}^{2}}.
\end{equation}

Therefore, all MUs experience the same SNR as expected, since the scaling factor is the same for all transmitted signals, and achieve the same rate 
\begin{equation}\label{eq:28}
R_{k}^{(\textrm{ERP})} = \log_{2}\left(1 + \gamma_{k}^{(\textrm{ERP})}\right). 
\end{equation}

Hence, the sum-rate of this ZF precoding scheme is given by:\begin{subequations}\label{eq:29}
\begin{align}
R_{\textrm{ERP}} &= \sum_{k=1}^{K}R_{k}^{(\textrm{ERP})}   \nonumber \\
&= \sum_{k=1}^{K}\log_{2}\left(1+\gamma_{k}^{\textrm{(ERP)}}\right)  \nonumber \\
&= \sum_{k=1}^{K}\log_{2}\left(1+ \frac{\beta}{\sigma_{n}^{2}}\right)  \nonumber \\
&= K\log_{2}\left(1 + \frac{\beta}{\sigma_{n}^{2}}\right) \label{eq:29a} \\
&= K\log_{2}\left(1 + \frac{1}{\sigma_{n}^{2}\sum_{k=1}^{K}1/\left(\lambda_{k}^{2}\right)}\right) \label{eq:29b}                                                                                      .
\end{align}
\end{subequations}

Another commonly used normalization, which results in higher sum-rate than ERP normalization, is the so-called equal transmit power (ETP) normalization which is obtained by setting
\begin{equation}\label{eq:30}
\matr{F} = \matr{H}^{+}= \matr{H}^{\dagger}\left(\matr{H}\matr{H}^{\dagger}\right)^{-1}
\end{equation}
(or $\matr{F} = \matr{H}^{-1}$ in case of a square channel matrix) and then dividing the elements of column $k$ of $\matr{F}$ with the norm of the corresponding column vector~\cite{ZFNorm}:
 \begin{equation}\label{eq:31}
\matr{W}_{\textrm{ETP}} = \frac{\matr{F}(:, k)}{\left\|\matr{F}(:,k)\right\|},\quad k = 1, \dots, K. 
 \end{equation}

The SINR in this case is
\begin{equation}\label{eq:32}
\gamma_{k}^{\textrm{(ETP)}} = \frac{1}{\sigma_{n}^{2}\left\|\matr{F}(:,k)\right\|^{2}}
\end{equation}
and the sum-rate is given as
\begin{align}
R_{\textrm{ETP}} &= \sum_{k=1}^{K}\log_{2}\left(1+\gamma_{k}^{\textrm{(ETP)}}\right)  \nonumber \\
&= \sum_{k=1}^{K}\log_{2}\left(1 +  \frac{1}{\sigma_{n}^{2}\left\|\matr{F}(:,k)\right\|^{2}}\right) \label{eq:33}.
\end{align}
\begin{figure*}[!t]
\centering
\includegraphics[scale = 0.4]{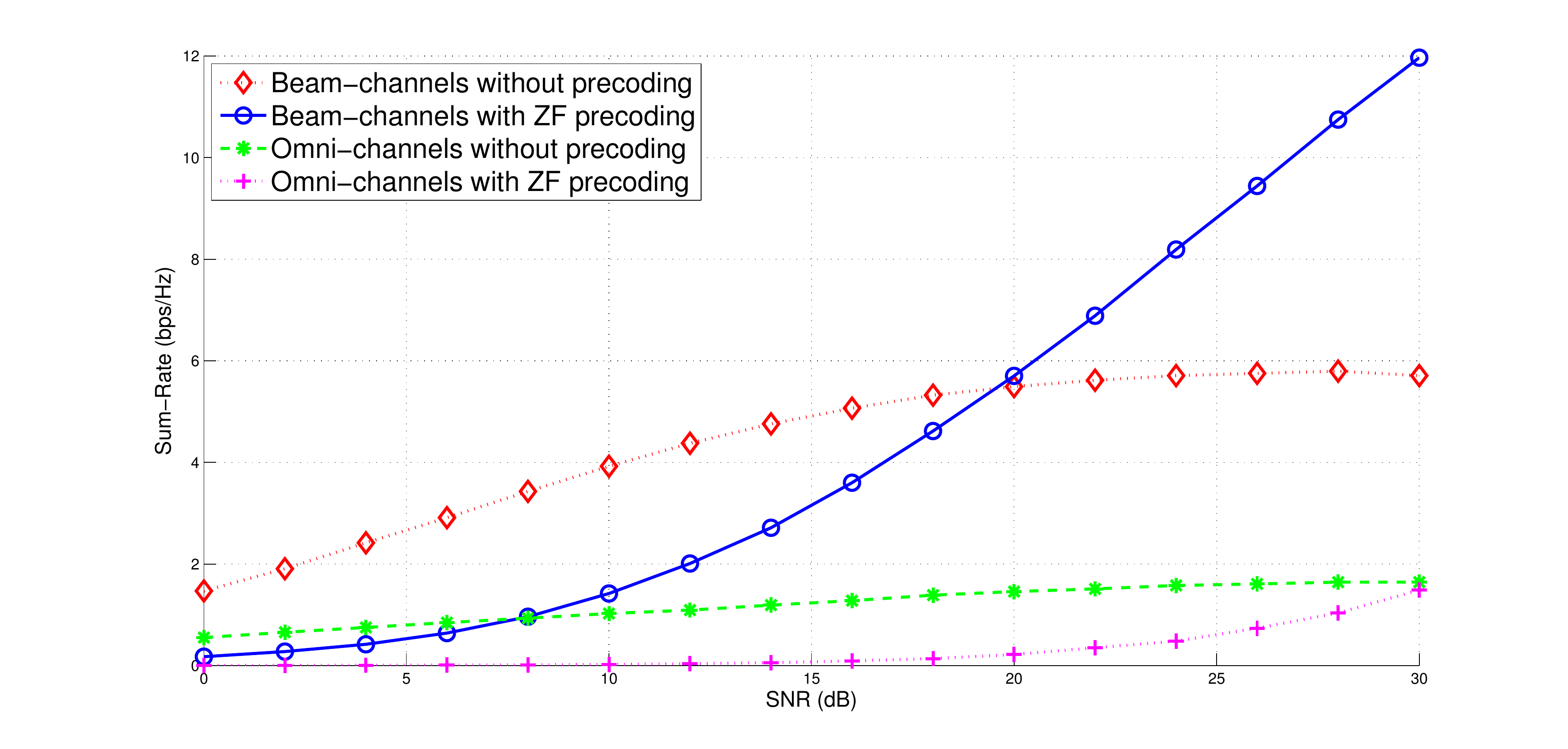}
\caption{Sum-rate of different transmission schemes for the case of $K = 2$ interfering links.}
\label{fig:4}
\end{figure*}
Both these normalization methods set MUI to zero and have the same total transmit power. These ZF precoding design schemes aim at minimizing transmit power. Another design approach is to set
\begin{equation}\label{eq:34}
\matr{W}_{\textrm{WF}} = \matr{H}^{\dagger}\left(\matr{H}\matr{H}^{\dagger}\right)^{-1}\textrm{diag}\left(\sqrt{P_{1}}, \dots, \sqrt{P_{K}}\right)
\end{equation}
and perform water-filling (WF) to optimally allocate transmit power per antenna $\left\{P_{k}\right\}_{k=1}^{K}$ such that the sum-rate is maximized, under the total power constraint
\begin{equation}\label{eq:35}
\sum_{k=1}^{K}\left\|\matr{w}_{k}\right\|P_{k} = P.
\end{equation}

However, this optimal ZF method is not applicable in the considered system setup due to the fact that it does not only increase the complexity of the cooperation between the BSs, but it requires also cooperation between the UEs (i.e., joint decoding).

\section{Performance Results}\label{sec:5}
In this section we evaluate the sum-rate performance of the proposed radio protocol via numerical simulations for the system setup described in Section~\ref{sec:3} with $K = 2$ and $L = 4$.

The performance results in Fig.~\ref{fig:4} represent ergodic sum-rates achieved over a range of target SNR values at the receiver, from 0dB up to 30dB. These results have been obtained after 1,000 simulation runs by taking the expectation of the corresponding sum-rate equations. Also, in each simulation run, 100 sub-runs are used for the normalization of the channel matrices, as described in Section~\ref{subsec:3.2}.

We consider transmission over the ``beam-domain'' channel with and without ZF precoding. That is, we study non-precoded transmission over the beams (which have been selected according to the SINRs that the MTs have fed back to the BSs) as well as the case where the MTs feed back the complex coefficients of the equivalent $2 \times 2$ ``beam-domain'' channel and ZF precoding is incorporated in the transmission process. 

For comparison purposes, we illustrate also the performance of the system when communication does not take place over a selected pair of beams, but we have instead conventional (non-precoded or precoded) ``omni-channel'' transmission. Of course, this type of non-precoded transmission would not take place in practice, since it is not a multi-user communication scheme, i.e., it does not take into account MUI; it is just included in the simulation as a lower performance bound. 

In our simulations we have set, for mathematical tractability and without loss of generality, $P = K$.

We note that non-precoded transmission over the ``beam-channels'' outperforms significantly ``omni-channel'' transmission methods, even when the latter employ ZF precoding, especially at high SNR values. This is due to: (\emph{a}) The gain that is introduced to the system as a result of the use of beams. (\emph{b}) The fact that the TXs are informed about the SINR at the RXs and select the best beam combination each time. 

We also note that the knowledge of the CSI at the transmitter for the equivalent ``beam-domain transformed'' channel and the joint precoding of the transmit vector that is performed based on that CSIT further improves the performance of the system at the high SNR regime. 

Finally, it is worth noting that the sum-rate of the transmission schemes that incorporate ZF precoding keeps increasing linearly within the considered SNR range, whereas the non-precoded techniques experience an expected flooring of their performance due to the residual interference that they incur, which is not accounted for.

\section{Complexity-Overhead and Complexity-Performance Tradeoffs}\label{sec:6}
In Section~\ref{sec:2}, a brief overview of ESPAR and C-RAN technologies was given. In this Section, we present the complexity reduction accomplished by the use of the proposed radio transmission scheme described in Section~\ref{sec:4}, mainly in terms of channel estimation / channel information feedback, and the relevant complexity-overhead as well as complexity-performance tradeoffs. But first, let us summarize the main points of Section~\ref{sec:2}:
\begin{itemize}
\item{ESPAR antennas reduce the size, complexity and cost of antenna arrays due to the fact that they use fewer active antenna elements and RF units than conventional antenna solutions and thus can be viewed as an enabler of massive MIMO.}
\item{C-RAN facilitates the efficient cooperation between individual BSs in Coordinated Multi-Point (CoMP) setups since the centralization and virtualization of the BBUs as well as the use of optical fibers at the fronthaul for the connection of the BBUs with the RRHs is capable of meeting the stringent delay and bandwidth requirements that are imposed by this family of cooperative MIMO communication techniques.}
\end{itemize}
The interested reader is encouraged to refer to the corresponding references cited in Section~\ref{sec:2} for a more detailed presentation of the benefits and drawbacks / challenges of these technologies.

As we mentioned in Section~\ref{sec:3}, our focus is on the downlink transmission in a system setup with $K$ cloud-based RRHs/BSs, each equipped with a single-RF ESPAR, and $K$ single-antenna UEs. Since the $K$ RRHs can be viewed as a composite BS with $K$ antennas due to the C-RAN architecture, MU-MISO communication techniques over a BC are applicable. 

The use of ESPARs instead of conventional antenna arrays reduces significantly channel estimation complexity due to the decreased number of active antenna elements which is translated into a correspondingly decreased number of direct- and cross-channels. In the remaining Section, we will compare ``omni-channel'' and ``beam-channel'' transmission methods in terms of complexity assuming in both cases the use of ESPARs at the BSs, but we should keep in mind this important note, that is, that the replacement of antenna arrays by ESPARs already provides complexity reduction regarding channel estimation.

Initially, let us assume conventional transmission over ``omni-channels''. The optimum precoding technique in terms of sum-rate is dirty paper coding (DPC)~\cite{MIMO}. However, the computational complexity of this non-linear precoding scheme makes it impractical. ZF precoding, on the other hand, is a low-overhead scheme that completely eliminates MUI, but it amplifies noise power. Minimum mean square error (MMSE) considers noise in the precoding design process, leading to better performance than ZF in low SNR conditions, at the cost of increased complexity~\cite{MIMO}. Hence, ZF precoding is the MU-MISO transmission method with the smaller overhead.

ZF precoding requires full CSIT. Since we have $K$ active antennas at the composite TX (and recalling that we are considering narrowband channels), the equivalent channel is described by a $K \times K$ matrix. In other words, each one of the $K$ UEs should feed back to its corresponding BS a vector of size $K$ containing the coefficient of the direct channel with that BS and the $K-1$ coefficients of the cross-channels with the other $K-1$ BSs. 

If we assume non-precoded transmission, as a low complexity bound, then each UE should feed back to its BS only its SINR. Thus, the composite BS would have to collect $K$ SINR values (real numbers), as opposed to the $K^{2}$ channel coefficients (complex numbers) required in the ZF precoding scenario. Of course, as we mentioned in the previous Section, such a transmission scheme would not be used in practice due to its poor performance.

Now, let us consider the proposed radio protocol. Each one of the $K$ RRHs is able to generate $L$ distinct beams. Thus, there exist $L^{K}$ possible beam combinations. Assuming non-precoded transmission (i.e., the beams to be used during transmission are selected according to criterion \#1 described in Section~\ref{sec:4}), each UE should feed back to its BS its SINR for each one of the beam combinations. That is, each one of the $K$ UEs should send to its corresponding BS $L^{K}$ SINR values. Therefore, $K\left(L^{K}\right)$ real values will be sent back to the composite BS in total. In contrast, when ZF precoding is used over ``omni-channels'', $K^{2}$ complex values have to be send back to the composite BS. However, in the latter case, accurate channel estimation is challenging when the number of BSs  (and active TX antenna elements) is high, while the measurement of the received SINR value at each UE is a rather trivial task. Also, while non-precoded, SINR-based transmission over ``omni-channels'' presents the lowest overhead (the feedback of only $K$ real values is required), it is not suitable for multi-user communication as we have already mentioned.

Let us turn our attention into ZF precoding-based transmission over the ``beam-channels'' (i.e., beam selection criterion \#2 is applied). For each beam-combination, the required CSIT is represented by a $K \times K$ complex-valued matrix. Thus, in total $K^{2}\left(L^{K}\right)$ complex values have to be fed back to the composite BS instead of only $K^{2}$ complex values that are required in the corresponding ``omni-channels'' case. However, a performance-complexity tradeoff is expected due to the ``beam-gains'' and the selection of the optimal ``composite beam-domain channel'', as we have seen in Section~\ref{sec:5}.  

Let us summarize: Non-precoded and precoded transmission over ``beam-channels'' results in an increase of the feedback overhead by $L^{K}$ in comparison with the corresponding ``omni-channel'' schemes due to the fact that each UE should send back to its BS its SINR or channel matrix \emph{for each beam combination}. However, it outperforms transmission over ``omni-channels'' and it simplifies (or even eliminates the need for) channel estimation. More specifically:
\begin{itemize}
\item{``Beam-channel'' communication techniques present a significant performance gain against their ``omni-channel'' counterparts.}
\item{Even non-precoded ``beam-channel'' transmission outperforms ZF ``omni-channel'' transmission. Thus, by using predetermined beams, we could simply collect SINR values instead of estimating channel matrices, which is a much simpler procedure - and therefore, it is easier to be used in practice, even in large setups.\footnote{In high mobility environments, it is possible to update the SINR value corresponding to each beam combination every time it is used to serve a user. Then, a record of time-windowed SINR values can be kept, thus further reducing the feedback requirements~\cite{Tse}\cite{JOBS}\cite{EUSIPCO}.}}
\item{The increase in the channel information overhead caused by the use of beams can be partly compensated by the use of ESPARs instead of conventional antenna arrays, depending on the ratio ``reduction in the number of active antenna elements / increase in the number of beams''}.
\end{itemize} 

\section{ZF Precoding-based Transmission over ``Beam-Channels'' with Imperfect CSIT}\label{sec:7}
So far, we have assumed perfect CSIT. In practice, though, various sources of error (e.g., channel estimation errors, channel quantization errors, feedback errors etc.) may result in imperfect CSIT. 

In this case, the imperfect channel matrix that is fed back to the TX is expressed as
\begin{equation}\label{eq:36}
\matr{H}_{e} = \matr{H} + \matr{E},
\end{equation}
where $\matr{H}$ is the actual channel matrix and $\matr{E}$ is an additive error matrix whose entries are independent and identically distributed (i.i.d.) RVs that are independent of $\matr{H}$ and follow a $\left[\matr{E}_{ij}\right]\sim \mathcal{CN}\left(0,\sigma_{e}^{2}\right)$ distribution.

Then, the precoded signal vector is expressed as
\begin{equation}\label{eq:37}
\matr{s}^{\prime} = \sqrt{\beta}\matr{W}_{\textrm{ZF}}\matr{H}_{e}^{+},
\end{equation}
where the power normalization factor $\beta$ is given by
\begin{subequations}\label{eq:38}
\begin{align}
\beta &= \frac{1}{\textrm{Tr}\left(\left(\matr{H}_{e}\matr{H}_{e}^{\dagger}\right)^{-1}\right)} \label{eq:38a}\\
&= \frac{1}{\left\|\left(\matr{H}_{e}\matr{H}_{e}^{\dagger}\right)^{-1}\right\|_{F}^{2}} \label{eq:38b}\\
&=  \frac{1}{\sum_{k=1}^{K}1/\left(\lambda_{k}^{2}\right)}, \label{eq:38c}
\end{align}
\end{subequations}  
and $\lambda_{k}$ is the $k$th singular value of $\matr{H}_{e}$.

The received signal vector is expressed as
\begin{align}
\matr{y} &= \matr{H}\matr{s}^{\prime} + \matr{n} \nonumber \\
&= \sqrt{\beta}\matr{H}\matr{H}_{e}^{\dagger}\matr{s}+ \matr{n} \nonumber \\
&= \sqrt{\beta}\left(\matr{H}_{e} - \matr{E}\right)\matr{H}_{e}^{\dagger}\matr{s}+ \matr{n} \nonumber \\
&= \sqrt{\beta}\matr{s} - \sqrt{\beta}\matr{E}\matr{H}_{e}^{\dagger}\matr{s}+ \matr{n}. \label{eq:39}
\end{align}

The SINR of each user, given that $\sigma_{s}^{2} = 1$, is
\begin{equation}\label{eq:40}
\gamma_{k}^{(\textrm{ZF})} = \frac{\beta}{P\sigma_{e}^{2} + \sigma_{n}^{2}} = \frac{1}{\left(\sigma_{e}^{2} + 1/P\right)\sum_{k=1}^{K}1/\lambda_{k}^{2}},
\end{equation}
while the sum-rate is expressed as
\begin{equation}\label{eq:41}
R_{\textrm{ZF}} = K\log_{2}\left(1+\frac{1}{\left(\sigma_{e}^{2} + 1/P\right)\sum_{k=1}^{K}1/\lambda_{k}^{2}}\right).
\end{equation}

Note that the variance of the error, $\sigma_{e}^{2}$, limits the sum-rate, i.e., as $P \rightarrow \infty$, 
\begin{equation}\label{eq:42}
R_{\textrm{ZF}} = K\log_{2}\left(1+\frac{1}{\left(\sigma_{e}^{2}\right)\sum_{k=1}^{K}1/\lambda_{k}^{2}}\right).
\end{equation}

We conclude this Section by evaluating the sum-rate performance of ZF precoding transmission over pre-determined beams for the considered setup in the case where CSIT is imperfect and for a range of error variance values $P_{e}$ from $10^{-3}$ up to $10^{-1}$ via simulations and we compare these results with the performance obtained when CSIT was assumed to be perfect.

\begin{figure*}[!t]
\centering
\includegraphics[scale = 0.4]{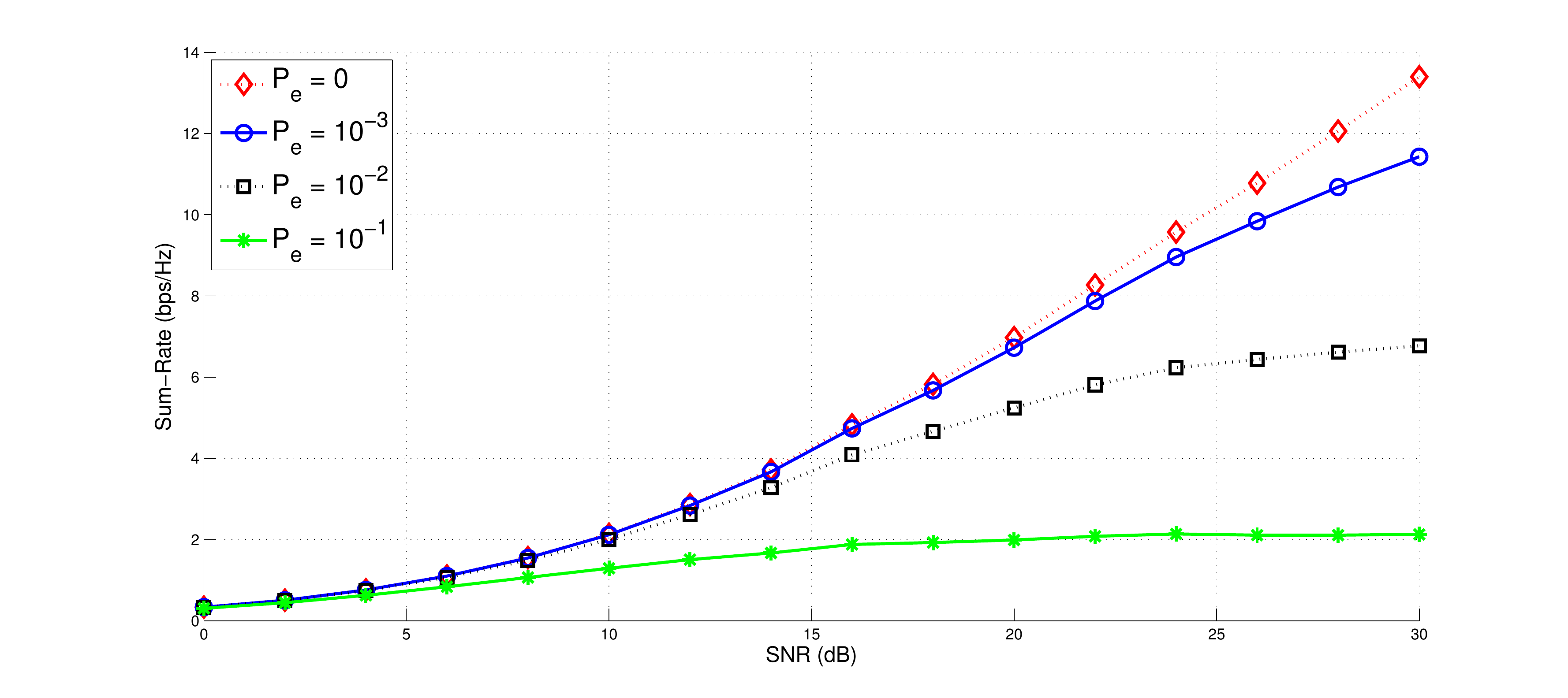}
\caption{Sum-rate of ZF precoding-based transmission over ``beam-channels'' for the case of imperfect CSIT and for various values of error variance.}
\label{fig:5}
\end{figure*}

As we see in Fig.~\ref{fig:5}, when the error variance is small, the degradation in the sum-rate throughput is negligible and only noticeable at the high SNR regime. However, as the error variance increases, the sum-rate is decreased and at some point floors (the higher the error variance, the sooner this flooring takes place). 

\section{Conclusions}\label{sec:8}
In this paper, we proposed a new radio transmission protocol for a system setup that could be an enabler of next-generation MU-MIMO systems. The proposed scheme performs extremely well in terms of the achieved sum-rate while it reduces significantly the complexity of the system. In the future, we plan to extend this work by studying larger setups and cases with imperfect CSI feedback.

\section*{Acknowledgments}
This work has been supported by the EC FP7 project HARP (http://www.fp7-harp.eu/) under grant number 318489.

\bibliographystyle{IEEEtran}
\bibliography{bibliography}
\end{document}